\documentclass[aip,reprint]{revtex4-1}

\usepackage{graphicx}
\usepackage{dcolumn}
\usepackage{bm}
\usepackage{amsmath}
\usepackage{xr}
\makeatletter
\newcommand*{\addFileDependency}[1]{
  \typeout{(#1)}
  \@addtofilelist{#1}
  \IfFileExists{#1}{}{\typeout{No file #1.}}
}
\makeatother

\newcommand*{\myexternaldocument}[1]{%
    \externaldocument{#1}%
    \addFileDependency{#1.tex}%
    \addFileDependency{#1.aux}%
}
\myexternaldocument{supplemental}

\draft 

\begin{document}


\title{Formation of zero-field skyrmion arrays in asymmetric superlattices} 



\author{Maxwell Li}%
 \email{mpli@andrew.cmu.edu.}
 \affiliation{Department of Materials Science \& Engineering, Carnegie Mellon University,\\ Pittsburgh, PA 15213 USA.}

\author{Anish Rai}%
 \affiliation{Department of Physics and Astronomy/MINT Center, The University of Alabama,\\ Tuscaloosa, AL 35487, USA.}
\author{Ashok Pokhrel}%
 \affiliation{Department of Physics and Astronomy/MINT Center, The University of Alabama,\\ Tuscaloosa, AL 35487, USA.}
\author{Arjun Sapkota}%
 \affiliation{Department of Physics and Astronomy/MINT Center, The University of Alabama,\\ Tuscaloosa, AL 35487, USA.}
\author{Claudia Mewes}%
 \affiliation{Department of Physics and Astronomy/MINT Center, The University of Alabama,\\ Tuscaloosa, AL 35487, USA.}
\author{Tim Mewes}%
 \affiliation{Department of Physics and Astronomy/MINT Center, The University of Alabama,\\ Tuscaloosa, AL 35487, USA.}

\author{Marc De Graef}%
 \affiliation{Department of Materials Science \& Engineering, Carnegie Mellon University,\\ Pittsburgh, PA 15213 USA.}
 \author{Vincent Sokalski}%
 \affiliation{Department of Materials Science \& Engineering, Carnegie Mellon University,\\ Pittsburgh, PA 15213 USA.}


\date{\today}

\begin{abstract}
 We demonstrate the formation of metastable N\'eel-type skyrmion arrays in Pt/Co/Ni/Ir multi-layers at zero-field following \textit{ex situ} application of an in-plane magnetic field using Lorentz transmission electron microscopy. The resultant skyrmion texture is found to depend on both the strength and misorientation of the applied field as well as the interfacial Dzyaloshinskii-Moriya interaction. To demonstrate the importance of the applied field angle, we leverage bend contours in the specimens which coincide with transition regions between skyrmion and labyrinth patterns. Subsequent application of a perpendicular magnetic field near these regions reveals the unusual situation where skyrmions with opposite magnetic polarities are stabilized in close proximity.
\end{abstract}


\maketitle 


Skyrmions are topologically-protected, particle-like objects that were first proposed by Skyrme\cite{Skyrme1962} and later theoretically predicted to be stabilized in chiral magnets\cite{Bogdanov1989,Bogdanov1994,Robler2006}. These magnetic skyrmions have since garnered a great deal of attention since their theoretical prediction and subsequent experimental discovery \cite{Muhlbauer2009,Yu2011,Fert2013,Jiang2015,Pollard2017,Everschor2018review,Jiang2019,Zhang2020review}. Properties such as topological protection, small size, and the high efficiency by which they can be manipulated with electric current make them promising candidates for use in future spintronic devices. Skyrmions with deterministic chirality are stabilized by the Dzyaloshinskii-Moriya interaction (DMI) which can be found in bulk magnetic materials lacking crystallographic inversion symmetry or in heavy metal/ferromagnet interfaces with large spin orbit coupling where z-mirror symmetry is also broken \cite{Dzyaloshinsky1958,Moriya1960,Thiaville2012}. Generally, skyrmions found in bulk materials have a Bloch-type configuration whereas those found in magnetic multi-layers have a N\'eel-type configuration consistent with C$_{\infty v}$ symmetry \cite{nagaosa2013}. In thin films, the application of a perpendicular field is generally required to stabilize skyrmions, which evolve reversibly from a labyrinth domain pattern as the field is increased \cite{Hubert1998}. 

In addition to understanding properties supporting their formation and controlling their size/mobility, methods for increasing the skyrmion density have also been investigated \cite{Yu2011,Zhang2018field,Zhang2018field}. Increased skyrmion density in thin films has been attributed to maximizing a critical parameter, $\kappa = \pi D/4\sqrt{AK_{\text{eff}}}$ where $D$ is DMI strength, which normalizes the reduction in the domain wall (DW) energy due to the DMI by the standard Bloch wall energy. Here, $A$ is exchange stiffness, and $K_{\text{eff}}$ is the effective magnetic anisotropy \cite{Bogdanov1994}. Previous works have demonstrated that temperature and in-plane fields assist in the formation of skyrmions \cite{Wang2017,Zhang2018field,Zhang2018temp,Qin2018}.  However, in most cases a perpendicular magnetic field was still required to prevent them from relaxing back into a labyrinth structure \cite{Wang2017,Zhang2018field,Zhang2018temp}. Field-free skyrmions have also been achieved through the application of voltage pulses, electric current pulses, or complex sample architecture \cite{Woo2016,He2017,Ma2019,Miao2014,Gilbert2015}. 

Historical work on uniaxial bubble materials considered the role of in-plane magnetic fields for breaking up labyrinth domain patterns, which we leverage to address open questions about the possible role of DMI in this process and its impact on skyrmion size and stability \cite{Kooy1960,Kaczer1961,Shimada1974}.  Here we study the formation of metastable skyrmion arrays at remanence after application of an \textit{ex situ} in-plane magnetic field. To vary important material parameters such as thickness, magnetic anisotropy, and DMI strength, we use a tunable asymmetric superlattice based on [Pt/(Co/Ni)$_M$/Ir]$_N$, where $M$ and $N$ can be independently varied \cite{Li2019}. The magnetic texture of these multi-layers is examined over a range of in-plane magnetic field magnitudes. Observations are compared to recent results on metastable skyrmions as well as those observed in magnetic bubble materials.

Multi-layers of [Pt(0.5nm)/(Co(0.2nm/Ni(0.6nm)$_M$/
Ir(0.5nm)]$_N$ were deposited onto $10$ nm thick amorphous Si$_3$N$_4$ membranes via magnetron sputtering in an Ar environment with a working pressure of $2.5$ mTorr and base pressure of $<3.0\times 10^{-7}$ Torr. An adhesion/seed layer of Ta(3nm)/Pt(3nm) and a Ta(3nm) cap are present in all films. Alternating gradient field magnetometry (AGFM) and vibrating sample magnetometry (VSM) were used to measure magnetic properties (saturation magnetization, $M_\text{S}$, and effective magnetic anisotropy $K_{\text{eff}}$). M-H loops confirm perpendicular magnetic anisotropy (PMA) in these multi-layers (Fig. S1). Although we find a monotonic increase of $K_{\text{eff}}$ with increased $M$, the observed $H_{\text{K}}$ values from these films remain very similar. Broadband ferromagentic resonance (FMR) spectroscopy was used to probe dynamic behavior and provides a more accurate measurement of the uniaxial magnetocrystalline anisotropy, $K_\text{u}$, which we find to be similar to the values calculated from M-H loops; these values as well as those calculated for damping constant, $\alpha$, are tabled in the Supplementary Information. Fresnel-mode LTEM images of [Pt/(Co/Ni)$_M$/Ir]$_N$ multi-layers were captured with an aberration-corrected FEI Titan G2 80-300 operated in Lorentz mode at room temperature \cite{degraef2000d}. A defocus value of -2.0 mm was used to capture the out-of-focus images shown here.

\begin{figure}[b]
\includegraphics{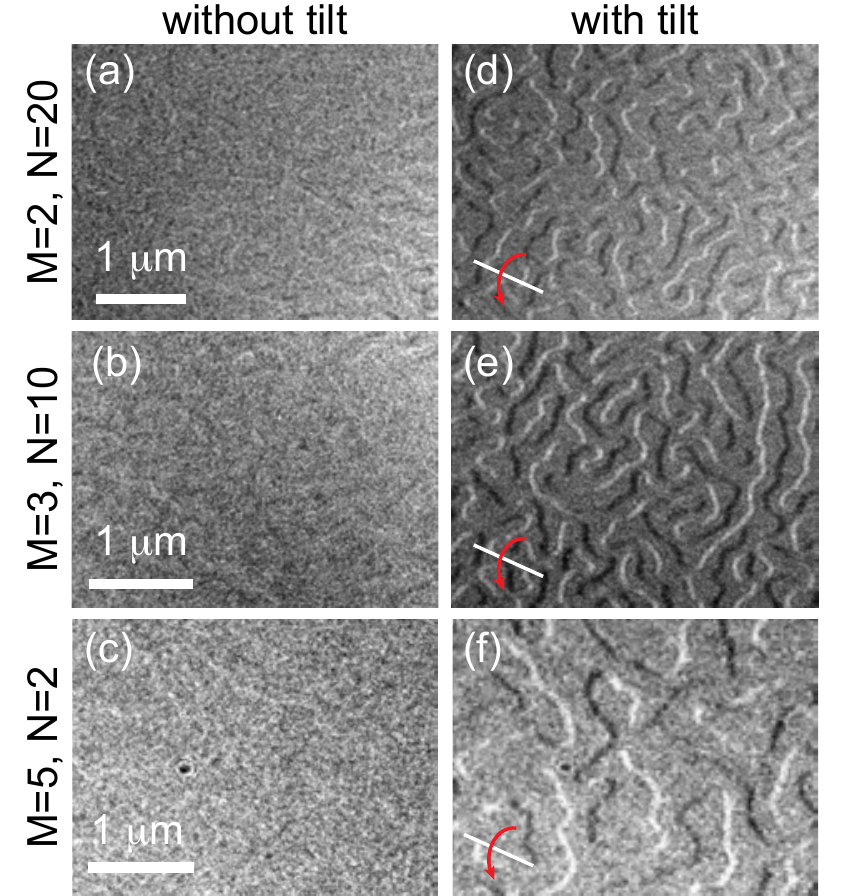}%
\caption{\label{labyrinths}Fresnel mode LTEM images of as-prepared [Pt/(Co/Ni)$_M$/Ir]$_N$ where (a,d) $M=2,N=20$, (b,e) $M=3,N=10$, and (c,f) $M=5,N=2$; magnetic contrast is only observed with the presence of sample tilt. The sample was tilted by 15$^\circ$ in (d-f).}%
\end{figure}

\begin{figure}[b]
\includegraphics{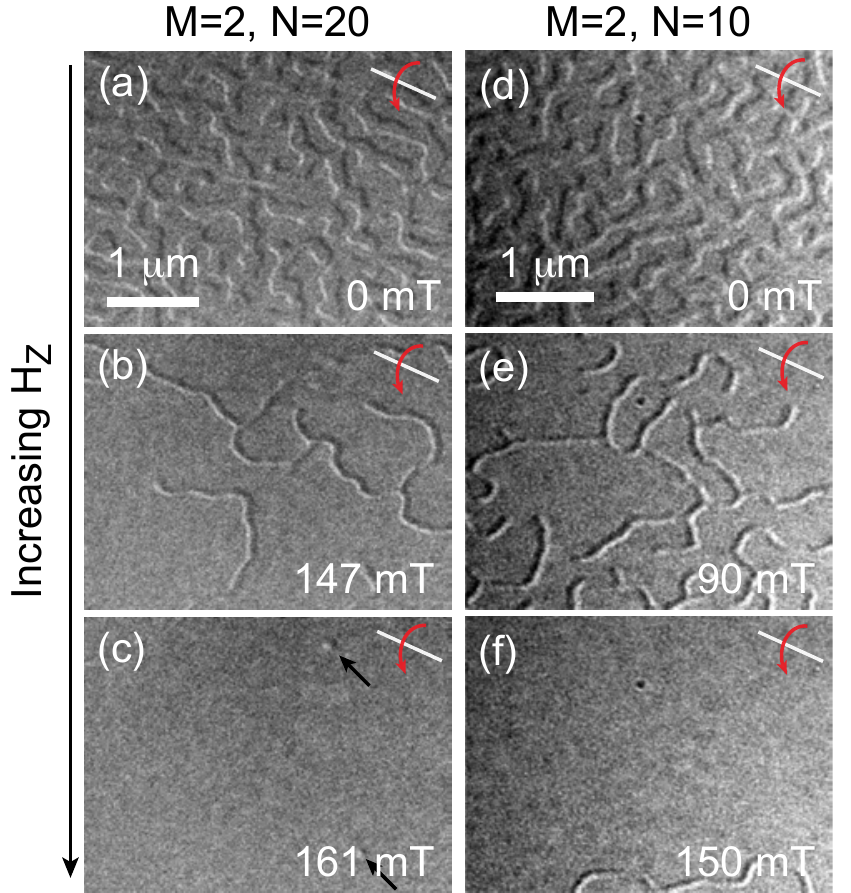}%
\caption{\label{Hz}Fresnel mode LTEM images of [Pt/(Co/Ni)$_M$/Ir]$_N$ where (a-c) $M=2,N=20$ and (d-f) $M=2,N=10$ in the presence of increasing perpendicular magnetic field applied \textit{in situ}. Isolated N\'eel skyrmions are highlighted with arrows in (c). A sample tilt of $15^\circ$ is present in each image.}%
\end{figure}

In the remnant state (after perpendicular saturation) for all samples investigated, magnetic contrast depicting a labyrinth domain structure is only observed with the application of sample tilt \textit{in situ} (Fig. \ref{labyrinths}). This indicates the presence of N\'eel walls and thus an appreciable interfacial DMI to stabilize them, which is expected based on direct measurements of DMI in this system \cite{Lau2018}. We now examine conventional skyrmion formation through application a perpendicular magnetic field \textit{in situ} by exciting the objective lens of the TEM. We note that in the presence of a sample tilt an effective in-plane field is generated. However, the magnitude of this field is not expected to have a significant effect of the resultant domain structure as it is much smaller than $H_\text{K}$. In a majority of these samples, instead of forming skyrmions, long worm-like domains form before annihilating at sufficiently large fields (Fig.~\ref{Hz} (d-f)); it is only in samples with large $N$, such as $M=2, N=20$, that skyrmions are observed to form in such a manner (Fig.~\ref{Hz} (a-c)).  This process is largely reversible as the labyrinth state is recovered after the field is removed.

\begin{figure*}[t]
\includegraphics{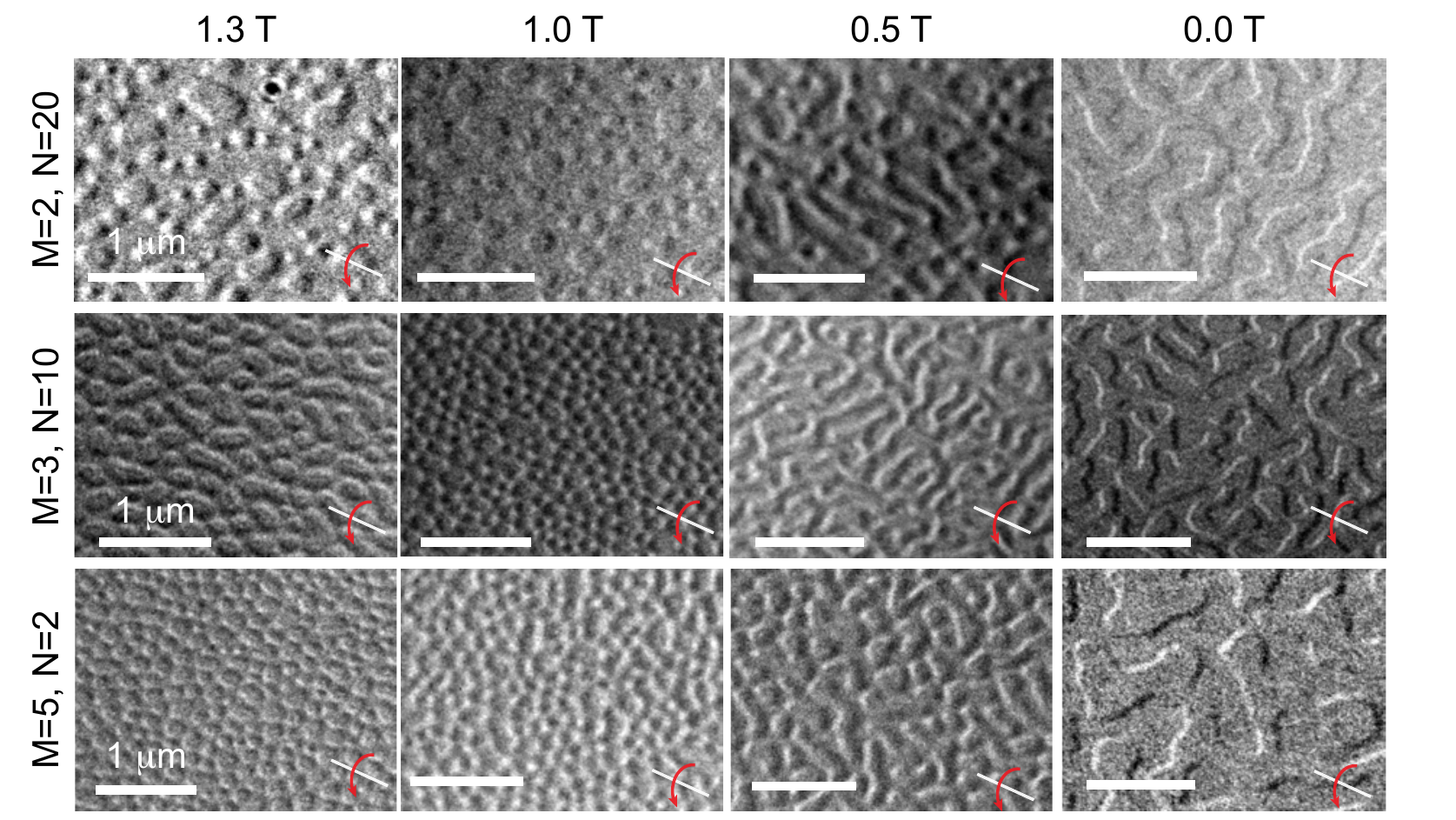}%
\caption{\label{arrays}Fresnel mode LTEM images of the remnant domain structure in [Pt/(Co/Ni)$_M$/Ir]$_N$ where top) $M=5,N=4$, middle) $M=3,N=10$, and bottom) $M=2,N=20$ following application of an in-plane magnetic fields of $1.3$, $1.0$, and $0.5$ T \textit{ex situ} and perpendicular saturation ($0.0$ T). The sample was tilted by $15^\circ$ in each image.}%
\end{figure*}

Next, in-plane magnetic fields are applied to these samples \textit{ex situ} before their remnant states are imaged with LTEM. Each sample was perpendicularly saturated and returned to remanence before in-plane fields were applied. Fresnel mode images reveal $\sim$100-200 nm diameter skyrmion arrays following the application of a $1.3$ T field (Fig.~\ref{arrays}). These arrays were also observed in samples that did not form skyrmions through the prior application of a perpendicular magnetic field. For the case of a $0.5$ T field, which is well-below the in-plane saturation field, Fresnel mode images revealed mixed stripe domains with some skyrmions, especially, in the higher DMI (M=2) case (Fig.~\ref{arrays}). The domain widths of these stripe domains are observed to be smaller than those of labyrinth domains in the demagnetized state (Fig. \ref{labyrinths}) and are comparable to those of field-free skyrmions. The observed skyrmion size from these micrographs was not clearly correlated with $H_\text{applied}$ beyond a critical value. However, skyrmion size was observed to decrease with decreasing $M$ (i.e. increasing DMI) which was not consistent with analytically calculated relationships between DMI strength and individual skyrmion size\cite{Wang2018}. However, because DMI decreases DW energy it can be reasoned that it suppresses the coarsening of individual skyrmions. In low DMI samples the system would tend towards coarsening in order to reduce the DW volume lending towards the observed larger sizes/smaller densities. This is in turn consistent with the notion that increasing $\kappa$ increases the resultant skyrmion density\cite{Zhang2018field}. Plots depicting observed skyrmion density and size with respect to $M$ can be found in the supplemental information. Further decreasing the applied field strengths reveal labyrinth domain structures at remnance. Our observations were compared with those from a symmetric [Co/Ni]$_{10.5}$ sample where DMI is expected to be negligible as confirmed by the observation of Bloch DWs (Fig. S7 (a))\cite{Li2019}. Following similar in-plane field treatment as the asymmetric samples, Fresnel mode LTEM images reveal the formation of field-free bubble arrays albeit with large numbers of vertical Bloch lines along their circumferences (Fig. S7 (b)); these bubbles have diameters that are notably larger than those observed in asymmetric samples with comparable thicknesses. This highlights the importance of DMI in reducing skyrmion size and controlling chirality in these field-free arrays.

It has previously been pointed out that there is a reduced contribution of $K_\text{eff}$ to DW energy upon application of in-plane fields, which serves to enhance the formation of skyrmions \cite{Zhang2018temp,Zhang2018field}. This helps explain the initial formation of skyrmions in response to an in-plane field, but it remains an open question why the skyrmion arrays observed here remain in the remnant state after its removal and why they appear to require a particularly large applied field to form. In the era of bubble materials, the formation of field-free arrays through the application of in-plane magnetic fields had also been examined in garnet films at much larger dimensions. \cite{Kooy1960,Kaczer1961,Shimada1974}. Shimada found that stripe domains form bubble arrays in the remnant state after application of an in-plane magnetic field due to the reduction of DW energy (similar to that noted previously) and consequent pinching to form bubble domains before they saturated \cite{Shimada1974}. The resultant metastable bubble domain formation is akin to grain coarsening or coalescence of soap bubbles, where growth rapidly decreases as the smallest features are absorbed into larger ones \cite{Babcock1989,Babcock1990}. In our case, we speculate that a combination of the energy barrier preventing annihilation of 360$^\circ$ transitions between skyrmions and pinning associated with our polycrystalline thin film microstructure counteract the driving force for the magnetic texture to return to a labyrinth ground state. This is supported by previous micromagnetic simulations where variations in local magnetocrystalline anisotropy, which simulates the disorder inherent to a polycrystalline film, promoted nucleation sites for skyrmions \cite{Zhang2018field,Wang2020}. Irrespective of the finer details of the formation mechanism, one key observation made by Shimada that we draw attention to in this work is that a slight misalignment of the applied field with respect to the film surface can greatly affect the resultant domain structure \cite{Kooy1960,Kaczer1961,Shimada1974}. 

\begin{figure}[t]
\includegraphics{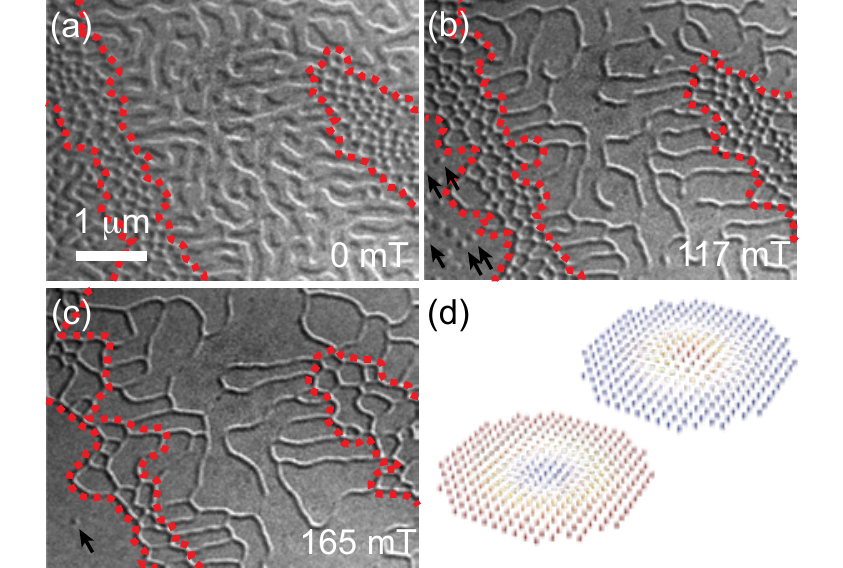}%
\caption{\label{bends}(a-c) Fresnel mode LTEM images of [Pt/(Co/Ni)$_3$/Ir]$_{10}$ with increasing perpendicular magnetic field applied \textit{in situ} at the location of a bend in the Si$_3$N$_4$ TEM membrane. Skyrmions of a polarity opposite that of the field direction are highlighted with black arrows whereas skyrmions with polarities parallel with the field direction are enclosed by a red dotted line. The sample was tilted by $15^\circ$ in each image. (d) Schematics depicting skyrmions with opposite polarities (e.g. blue points ``down", red points `up").}%
\end{figure}

To test the role of a misaligned in-plane magnetic field in our samples, we leverage bend contours associated with the TEM membranes.  Fallon, et al. have found that the flexible membrane can tilt as much as $15^\circ$, which is consistent with our observations\cite{Fallon2020}.  Throughout each sample, transition regions between a labyrinth domain structure and a skyrmion array (noted by red lines in Fig. \ref{bends}) were observed at the location of bends in the specimen.  Although it would otherwise be reasonable to speculate that this magnetic texture could result from magnetoelastic effects, the analyses provided by Shimada make clear that the origin is actually due to variations in the applied field angle with the sample surface\cite{Shimada1974}. The regions that retain a labyrinth pattern are those which were significantly offset from the applied field direction.  Subsequent \textit{in situ} application of a perpendicular magnetic field at these labyrinth regions did not lead to the formation of skyrmions despite being surrounded by skyrmion arrays (Fig. \ref{bends} (b-c)). Additionally, when the domain pattern of (Fig. \ref{bends} (a)) is examined during \textit{in situ} application of a perpendicular magnetic field, we find the unusual circumstance where skyrmions of both polarities exist (i.e. the magnetization at their cores point in opposite directions) in close proximity.  This is because the labyrinth regions enter the conventional transition to skyrmions while the metastable skyrmion arrays have not yet begun to coalesce. 

In summary, metastable skyrmion arrays were observed at the remnant state in [Pt/(Co/Ni)$_M$/Ir]$_N$ multi-layers after \textit{ex situ} application of a large in-plane magnetic field -- an approach adopted from prior work on bubble materials \cite{Shimada1974}. The average skyrmion size was not strongly correlated to the strength of the field applied beyond 1 T but appeared inversely related to the interfacial DMI strength. The metastability of these arrays likely originates from a combination of the DMI-reduced wall energy, which suppresses the driving force for coarsening, and the energy barrier associated with annihilation of 360$^\circ$ transitions between skyrmions. We also speculate that intrinsic pinning of the polycrystalline microstructure contributes to array formation. The origin of morphological transitions between labyrinths and skyrmion arrays throughout bent specimen membranes is confirmed to originate from variations in the applied field angle with respect to the film plane emphasizing the importance of such angle. Additionally, the simultaneous observation of skyrmions with opposite polarities was seen at these bends. This simple method of forming dense, field-free skyrmion arrays, coupled with our highly tunable Pt/Co/Ni/Ir system, offers a means of investigating skyrmions out of equilibrium for future development of spintronic devices.\\

\section*{\label{sec:level0}Supplementary Material}
Supplementary material contains detailed explanation of FMR measurements, values of measured magnetic parameters, plots detailing skyrmion size/density, and additional Lorentz TEM images.

\begin{acknowledgments}
This work is financially  supported by the Defense Advanced Research Project Agency (DARPA) program on Topological Excitations in Electronics (TEE) under grant number D18AP00011. The authors also acknowledge use of the Materials Characterization Facility at Carnegie Mellon University supported by grant MCF-677785. M.L. would like to also acknowledge Michael Kitcher for assistance with calculation of skyrmion schematics used.
\end{acknowledgments}

\section*{\label{sec:level0}Data Availability Statement}
The data that supports the findings of this study are available within the article and its supplementary material.

\bibliography{aipsamp}

\end{document}


\title{Supplementary material for ``Formation of zero-field skyrmion arrays in asymmetric superlattices"}

\author{Maxwell Li}%
 \email{mpli@andrew.cmu.edu.}
 \affiliation{Department of Materials Science \& Engineering, Carnegie Mellon University, Pittsburgh, PA 15213 USA.}
\author{Anish Rai}%
\author{Ashok Pokhrel}%
\author{Arjun Sapkota}%
\author{Claudia Mewes}%
\author{Tim Mewes}%
 \affiliation{Department of Physics and Astronomy/MINT Center, The University of Alabama, Tuscaloosa, AL 35487, USA.}

\author{Marc De Graef}%
\author{Vincent Sokalski}%
 \affiliation{Department of Materials Science \& Engineering, Carnegie Mellon University, Pittsburgh, PA 15213 USA.}

\maketitle

\begin{figure}[h]
\includegraphics{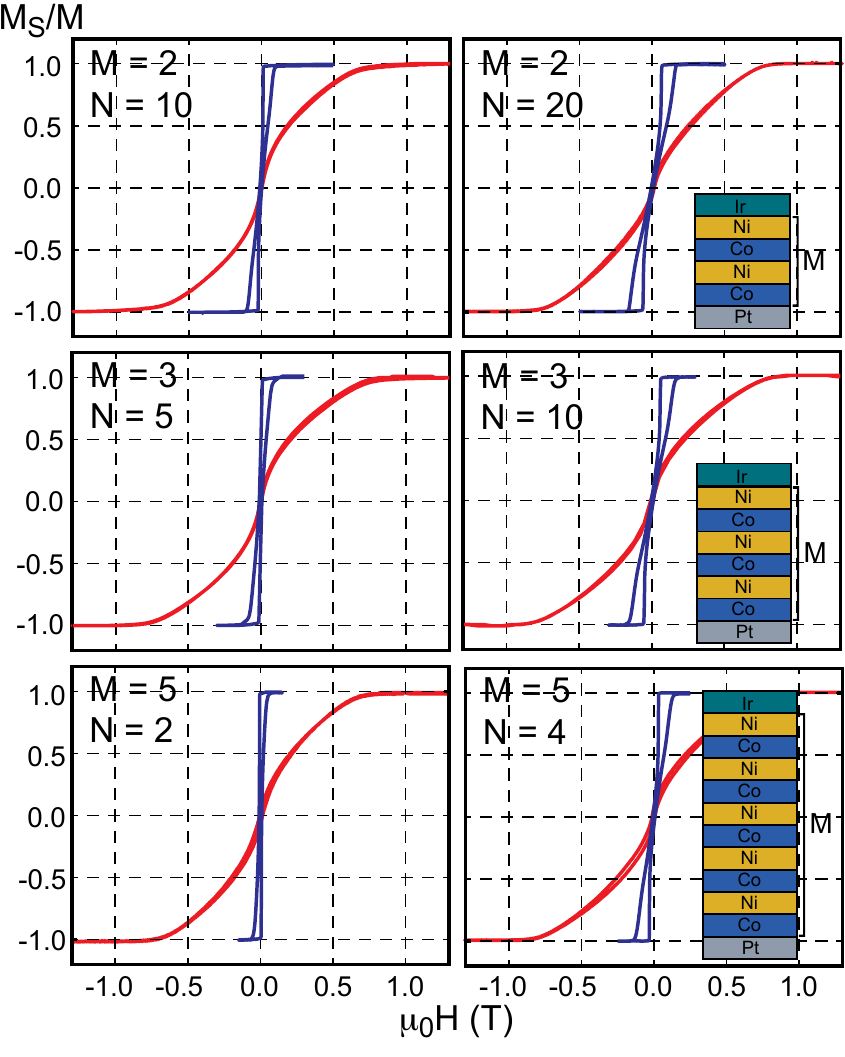}%
\caption{\label{MHloops}M-H loops of Pt/Co/Ni/Ir multi-layers examined in this work where blue and red represent perpendicular and parallel loops, respectively. Insets depict schematics of the multi-layer design.}%
\end{figure}

Saturation magnetization, $M_\text{S}$, and effective magnetic anisotropy, $K_{\text{eff}}$, of each sample examined were determined via vibrating sample mangetometry (VSM). The M-H loops from these measurements are depicted in Fig. \ref{MHloops}. The $K_{\text{eff}}$ is defined as the area between the perpendicular and parallel M-H loops. Both $M_{\text{S}}$ and $K_{\text{eff}}$ were observed to increase monotonically with $M$ consistent with observations in our previous work \cite{Li2019}.

\begin{figure}[h]
\includegraphics{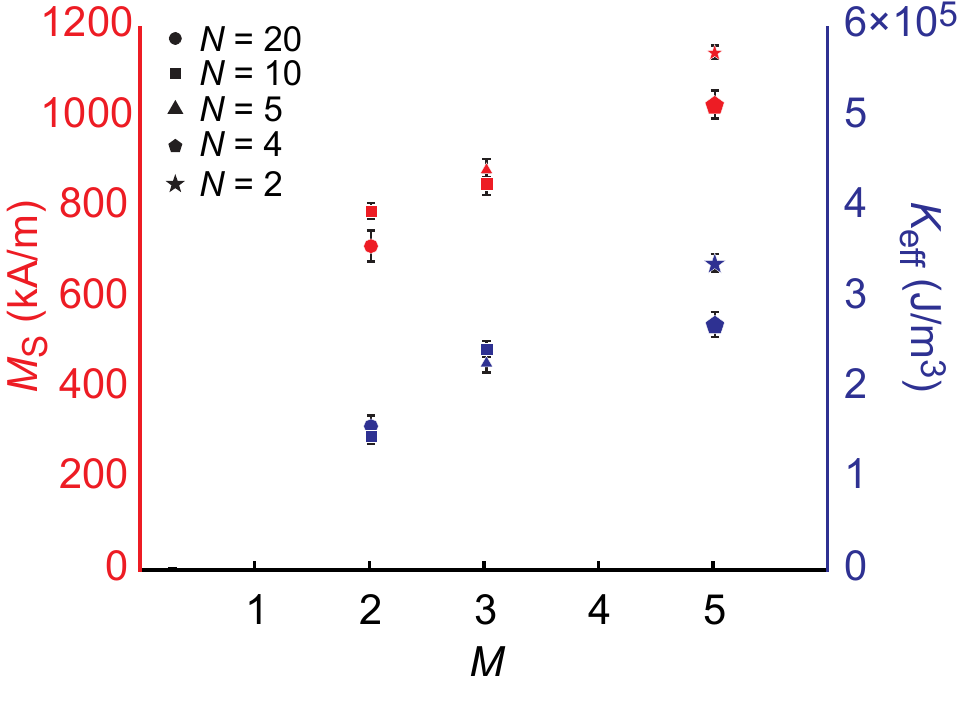}%
\caption{\label{VSM} $M_\text{S}$ and $K_\text{eff}$ values derived from VSM measurements for Pt/Co/Ni/Ir multi-layers examined in this work. $K_\text{eff}$ was derived from the area between the perpendicular and parallel M-H loops.}%
\end{figure}

\clearpage
Ferromagnetic resonance (FMR) measurements were carried out with a custom designed broadband coplanar waveguide room temperature setup. This setup enables the measurement of field-swept spectra at arbitrary frequencies up to 65 GHz. The instrument also allows to change the in-plane angle and out-of-plane of the applied field if needed. The recorded FMR spectra are analyzed by fitting the derivative of a Lorentzian at each frequency to determine the frequency dependence of the resonance field $H_{res}$ and resonance peak-to-peak linewidth $\Delta{H}$ \cite{oates2002high,pachauri2016comprehensive}.
For the samples discussed in this article broadband FMR measurements were carried out with the external magnetic field applied in-plane (ip) and out-of-plane (oop). These measurements can be utilized to extract the uniaxial anisotropy of the samples. For all the samples, there is a significant contribution of a higher order anisotropy, which is evident by a significant difference of the effective magnetizations $M_{\text{eff}}^{ip}$   and $M_{\text{eff}}^{oop}$ determined in the two configurations. For samples with both a second $\tilde{K}_{\mathrm{2,eff}}$ and fourth-order uniaxial anisotropy $K_{\mathrm{4,eff}}$ present the effective magnetization measured along the two orientations is given by \cite{shaw2015perpendicular,mohammadi2018inhomogeneous}:
\begin{equation}
\label{K2}
M_{\mathrm{eff}}^{\mathrm{ip}}=\frac{2\tilde{K}_{\mathrm{2,eff}}}{\mu_\mathrm{0}M_\mathrm{s}}
\end{equation}
\begin{equation}
\label{K4}
M_{\mathrm{eff}}^{\mathrm{oop}}=\frac{2\tilde{K}_{\mathrm{2,eff}}}{\mu_\mathrm{0}M_\mathrm{s}}+\frac{2K_{\mathrm{4,eff}}}{\mu_\mathrm{0}M_\mathrm{s}}
\end{equation}

$\tilde{K}_{\mathrm{2,eff}}$ contains both shape anisotropy and second order uniaxial anisotropy contribution $K_{2,\text{eff}}$. For infinite, homogeneously magnetized thin films, one can express $\tilde{K}_{\mathrm{2,eff}}$ as $\tilde{K}_{\mathrm{2,eff}}=\frac{\mu_\mathrm{0}M_\mathrm{s}}{2}+ K_{2,\text{eff}}$. For samples having just a second order magnetic anisotropy, one can express the anisotropy field as:
\begin{equation}
\mu_\mathrm{0}H_\mathrm{k} = \frac{2K_{2,\text{eff}}}{\mu_\mathrm{0}M_\mathrm{s}}
\end{equation}
With the presence of higher order magnetic anisotropy, one can express the anisotropy field as:
\begin{equation}
\mu_\mathrm{0}H_\mathrm{k} = \frac{2K_{2,\text{eff}}}{\mu_\mathrm{0}M_\mathrm{s}}+\frac{2K_{\mathrm{4,eff}}}{\mu_\mathrm{0}M_\mathrm{s}}
\end{equation}

\begin{figure}[h]
\includegraphics[width=0.6\textwidth]{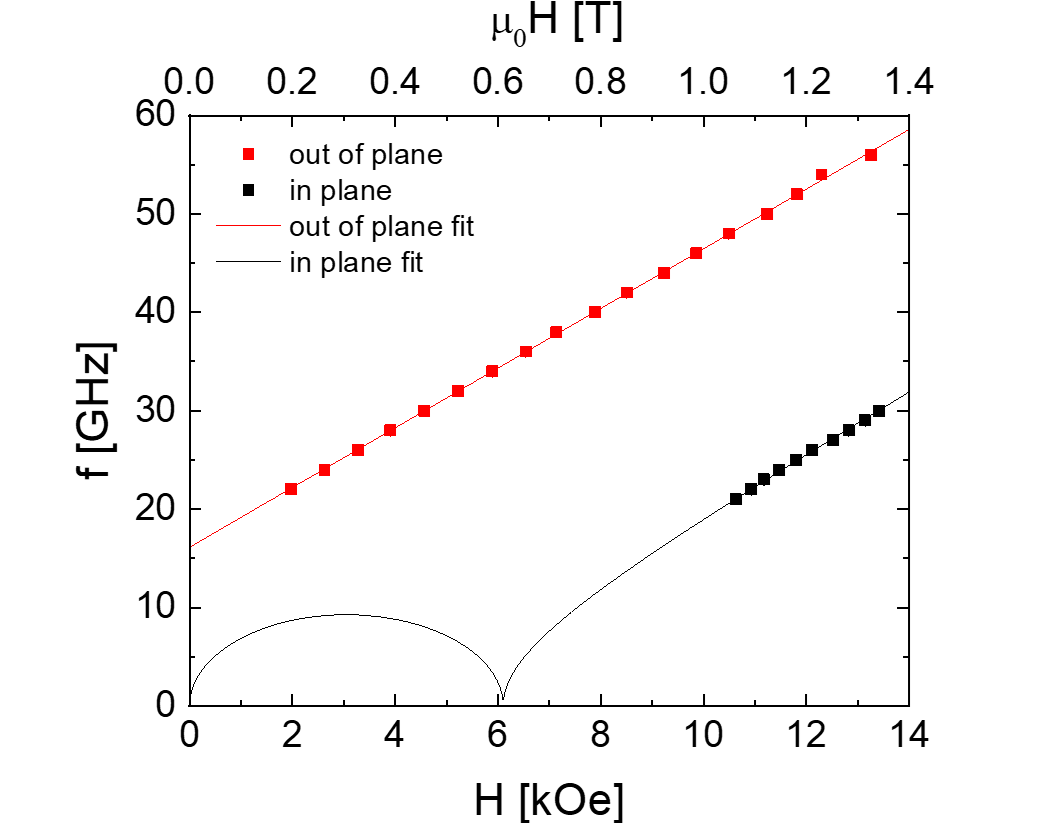}%
\caption{\label{KittelPlot}Kittel plots for $[\mathrm{Pt/(Co/Ni)}_{\mathrm{5}}/\mathrm{Ir}]_{\mathrm{2}}$ sample. The red squares and line represent out of plane data and fit and The red squares and line represent in plane data and fit respectively.}%
\end{figure}

The data in Fig. \ref{KittelPlot} are fitted with Kittel equations of the form 
\begin{equation}
\label{KittelParallel}
f^{\mathrm{ip}}={\gamma}^{'}\sqrt{H_{\mathrm{res}}\left(H_{\mathrm{res}}+M_{\mathrm{eff}}^{\mathrm{ip}}\right)}
\end{equation}
\begin{equation}
\label{KittelPerpendicular}
f^{\mathrm{oop}}={\gamma}^{'}\left(H_{\mathrm{res}}-M_{\mathrm{eff}}^{\mathrm{oop}}\right)
\end{equation}
Where $\gamma'$ is the reduced gyromagnetic ratio and is a shared parameter for fitting. From Fig. \ref{KittelPlot}, it is also clear that the easy axis of the sample is along the out-of-plane axis.\\\\

\begin{table}[h]
  \caption{\label{tab:table1} Measured values of $K_\text{eff}$ via VSM, and $K_2$, $K_4$, and $\alpha$ via FMR.}
  \begin{ruledtabular}
  \begin{tabular}{c|c||c c c c}
  $M$ & $N$  &   $K_{\text{eff}}$ (J/m$^3$)    &   $K_2$ (J/m$^3$),    &   $K_4$ (J/m$^3$)    &   $\alpha$\\
  \hline%
  5 & 2 & $1.50\times10^5$ & -1.55$\times10^5$ & -1.92$\times10^4$ & $0.0303\pm0.0008$\\
  3 & 10 & $2.43\times10^5$ & -2.59$\times10^5$ & 3.36$\times10^4$ & $0.057\pm0.003$\\
  2 & 20 & $3.37\times10^5$ & -3.55$\times10^5$ & 7.23$\times10^4$ & --- \\
 \end{tabular}
 \end{ruledtabular}
 \end{table}

\newpage
The observed skyrmion density and average size for the multi-layers examined in this work are plot in the figures below. Skyrmion density is generally observed to decrease with increasing $M$ which is likely related to a decreased DMI strength. A larger $D$ would be expected to promote smaller sized skyrmions and thus increase the areal density of skyrmions observed in a sample. This trend is supported by the average skyrmion diameter observed with respect to $M$ as seen below. 

\begin{figure}[h]
\includegraphics{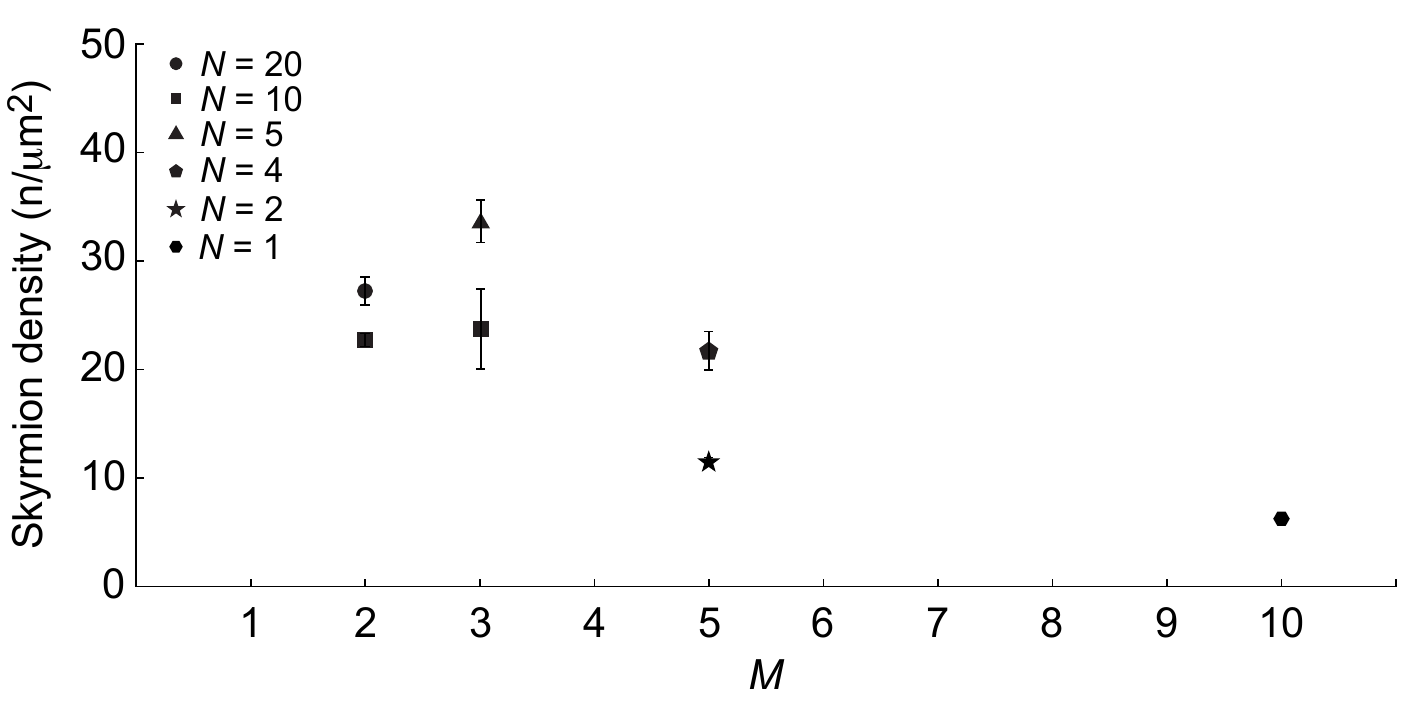}%
\caption{\label{density} Average skyrmion density with respect to ferromagnetic thickness, $M$.}%
\end{figure}

\begin{figure}[h]
\includegraphics{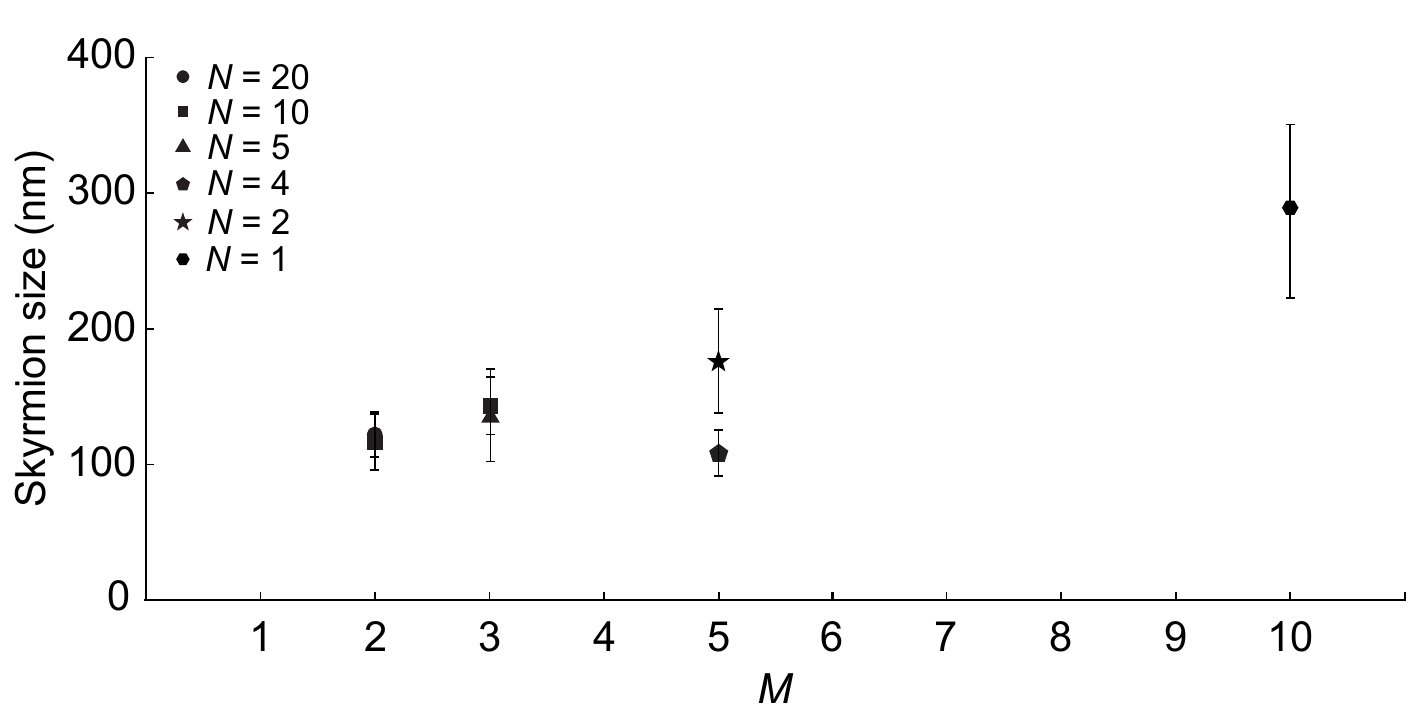}%
\caption{\label{size} Average skyrmion diameters found in arrays with respect to ferromagnetic thickness, $M$.}%
\end{figure}

\newpage
The average size of skyrmions was determined as the distance between the intensity maximum and minimum of a line scan across their width as described by Pollard, et al. \cite{Pollard2017}. It is seen that the average skyrmion diameter and distribution of measured diameters generally increases with decreasing DMI (via $M$) (Fig. \ref{size}).

\begin{figure}[h]
\includegraphics[width=0.8\textwidth]{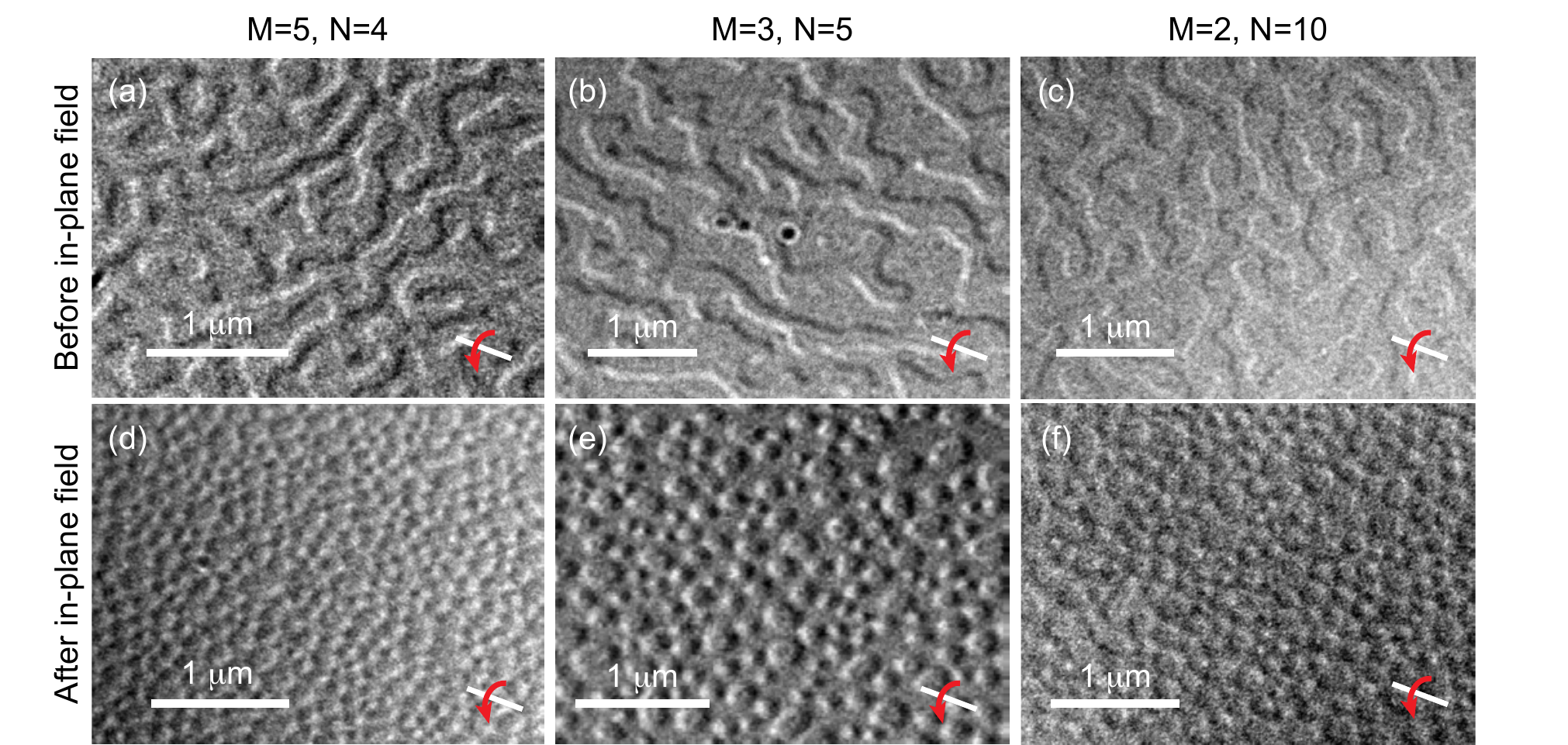}%
\caption{\label{saturating} Fresnel mode Lorentz TEM images of [Pt/(Co/Ni)$_M$/Ir]$_N$ multi-layers at the remnant state following perpendicular saturation (a-c) or application of a 1.3 T in-plane magnetic field applied \textit{ex situ} (d-f). The sample was tilted by $15^\circ$ in each image.}%
\end{figure}

\newpage
Additional Fresnel mode Lorentz TEM images of Pt/Co/Ni/Ir multi-layers (Fig. \ref{saturating}) depict remnant states of [Pt/(Co/Ni)$_M$/Ir]$_N$ films after perpendicular saturation (a-c) or application of a 1.3 T in-plane magnetic field (d-f). 

\newpage
Fresnel Lorentz TEM images of symmetric [Co/Ni]$_{10.5}$ samples which served as a limiting case where DMI strength is not expected to be significant. Bloch DWs, which readily display magnetic contrast in the absence of sample tilt, are observed in addition to a large number of vertical Bloch lines at the remnant state following perpendicular saturation. After an exposure to a large in-plane magnetic field, the remnant state reveals a magnetic bubble array similar to those observed in asymmetric samples albeit with a large presence of vertical Bloch lines. The lack of preferred chirality observed here is indicative of the negligible DMI strength in these symmetric films. Although DMI does not appear to influence the formation of these field-free bubble/skyrmion arrays, a significant DMI is still needed to form homochiral skyrmions.

\begin{figure}[h]
\includegraphics[width=0.8\textwidth]{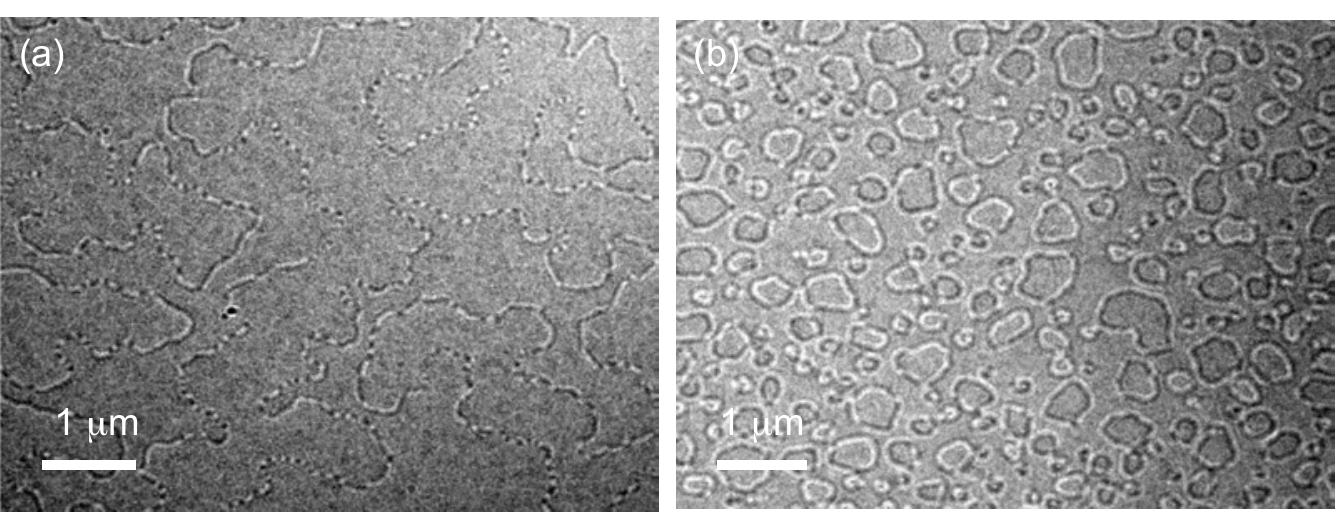}%
\caption{\label{SICoNi} Fresnel mode Lorentz TEM images of symmetric [Co/Ni]$_{10.5}$ multi-layers at the remnant state following (a) perpendicular saturation and (b) application of a 1.3 T in-plane magnetic field applied. Strong Bloch domain wall character indicates a negligible interfacial DMI.}%
\end{figure}

\begin{figure}[h]
\includegraphics[width=1.0\textwidth]{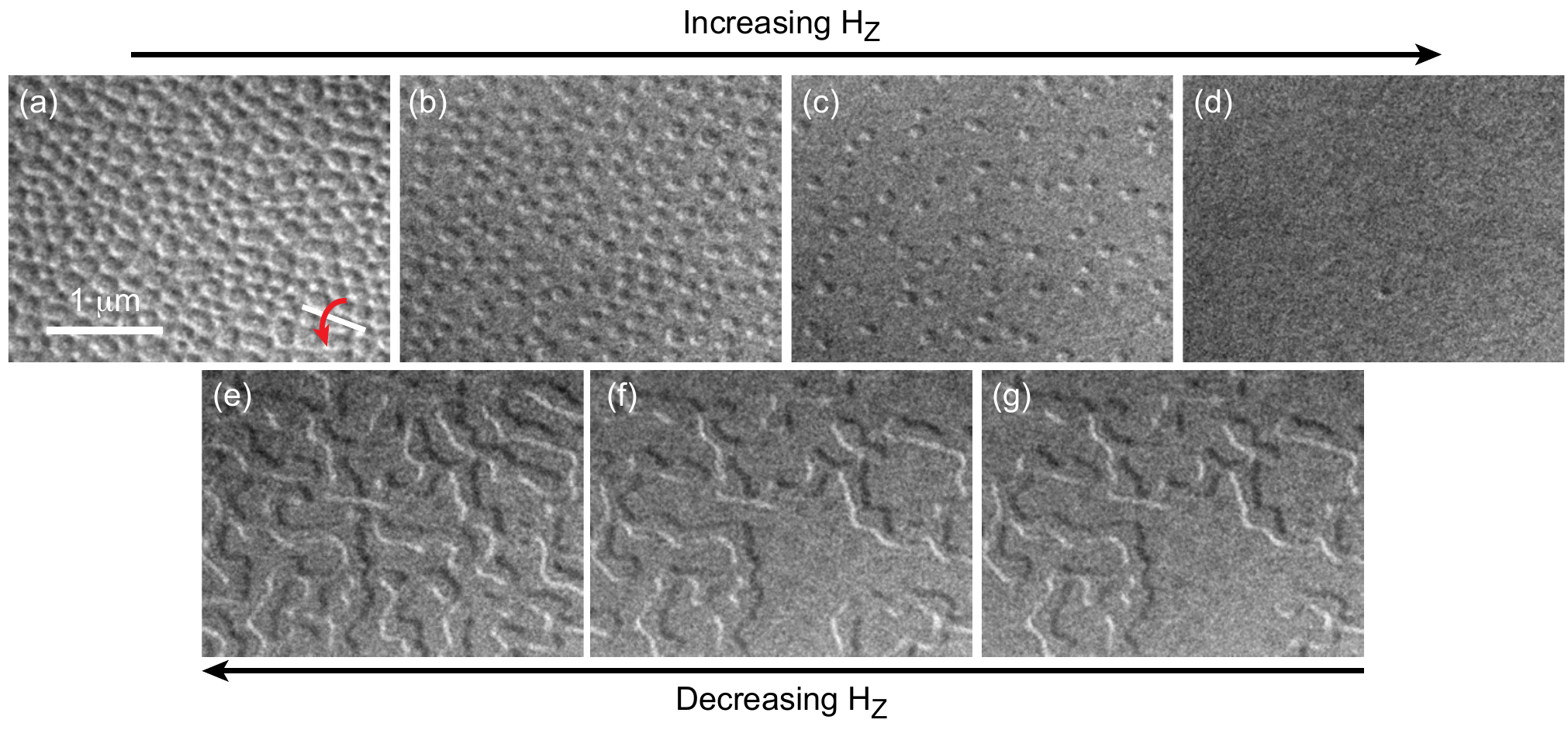}%
\caption{\label{labyrinthremnance} Fresnel mode Lorentz TEM images of a skyrmion array in [Pt/(Co/Ni)$_2$/Ir]$_{20}$ with varying perpendicular magnetic fields applied \textit{in situ}. (a-d) depict change in domain structure with increasing fields while (e-g) show the nucleation of domains following perpendicular saturation. The sample was tilted by $15^\circ$ in each image.}%
\end{figure}

\begin{figure}[h]
\includegraphics[width=1.0\textwidth]{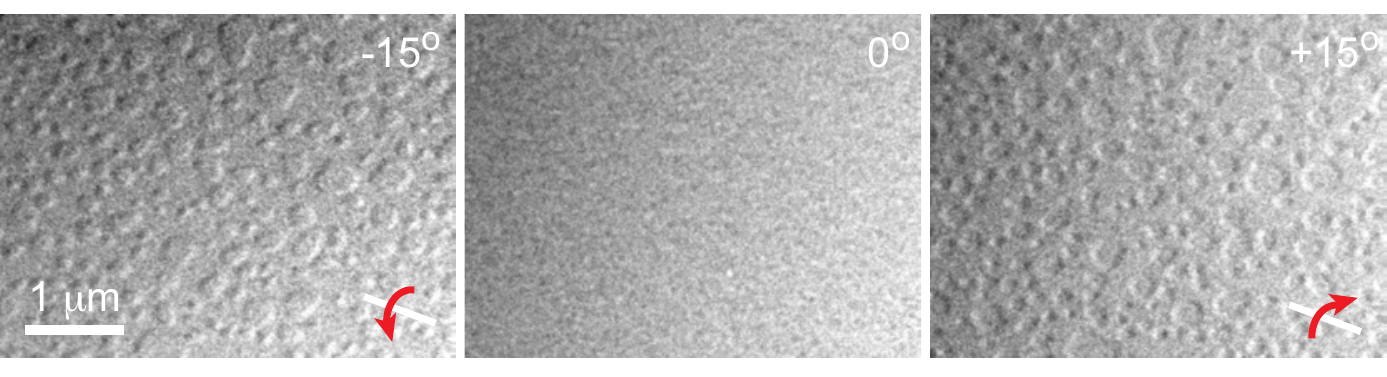}%
\caption{\label{SItilt} Fresnel mode Lorentz TEM images of a skyrmion array in [Pt/(Co/Ni)$_5$/Ir]$_{2}$ at various applied sample tilts of $-15^\circ$, $0^\circ$, and $+15^\circ$. It can be seen that the magnetic contrast disappears at zero tilt and inverts through the tilt series.}%
\end{figure}

\newpage
Additional Fresnel mode Lorentz TEM images of labyrinth domain formation after skyrmion arrays are brought to saturation via perpendicular magnetic field applied \textit{in situ} is also shown in Fig. \ref{labyrinthremnance}. Images of skyrmion arrays through a tilt series (Fig. \ref{SItilt}) show the DW character of films are still primarily N\'eel after skyrmion array formation as no magnetic contrast is observed at zero tilt. Additionally, magnetic contrast is observed to invert through the tilt series as it would in a through-focus series.  

\newpage
\bibliography{aipsamp}